\documentclass[12pt]{article}
\usepackage{amssymb}
\usepackage{amsfonts}
\usepackage{graphicx}
\usepackage{epsfig}

\newcommand{\ep}{{\cal E}_{||}}
\newcommand{\ec}{{\cal E}_{C}}

\newcommand{\epp}{${\cal E}_{||}$}
\newcommand{\ecc}{${\cal E}_{C}$}
\newcommand{\ess}{${\cal E}_{S}$}
\newcommand{\eppt}{${\cal E}^{(2)}_{||}$}
\newcommand{\eps}{\varepsilon}
\newcommand{\ecct}{${\cal E}^{(2)}_{C}$}
\newcommand{\esst}{${\cal E}^{(2)}_{S}$}
\newcommand{\eccf}{${\cal E}^{(4)}_{C}$}
\newcommand{\equ}[1]{\ Eq.~(\ref{#1})}
\newcommand{\ini}{\begin{equation}}
\newcommand{\bel}[1]{\begin{equation}\label{#1}}
\newcommand{\bal}[1]{\begin{eqnarray}\label{#1}}
\newcommand{\fin}{\end{equation}}
\newcommand{\ea}{\end{eqnarray}}
\newcommand{\grad}{\vec\nabla}
\newcommand{\tp}{^{(2)}_{||}}
\newcommand{\rr}{{\cal R}}
\newcommand{\qb}{\bar{q}}
\newcommand{\qbp}{\bar{q}'}
\newcommand{\pb}{\bar{p}}
\newcommand{\pbp}{\bar{p}'}
\newcommand{\cale}{{\cal E}}
\newcommand{\bmath}{\begin{displaymath}}
\newcommand{\emath}{\end{displaymath}}
\newcommand{\bite}{\begin{itemize}}
\newcommand{\eite}{\end{itemize}}
\newcommand{\bbeta}{\hbox{$\beta\hspace{-.645em}\beta\hspace{-.645em}\beta$}}
\newcommand{\bsbeta}{{\beta\hspace{-.46em}\beta\hspace{-.46em}\beta}}
\newcommand{\tw}{{(2)}}
\newcommand{\vn}{\vec{n}}

\newcommand{\calem}{{\cal E}^{EM}}
\newcommand{\mod}[1]{\ {\rm mod}\;#1}
\newcommand{\drop}[1]{}
\renewcommand{\today}{October 26, 2005}

\begin{document}

\vspace{0.5cm}
\begin{center}
{\Large \bf Comments on the Sign and Other Aspects of Semiclassical Casimir Energies}\\
\vspace{0.7cm}
\vspace{0.5cm}
{\large
 Martin Schaden}\\

\vspace{1.5cm} {\it\small Rutgers University, 365 Smith Hall, 101
Warren St., Newark, NJ 07102, USA.\vspace{1.5cm}}\\
\today
\end{center}

\bigskip
\begin{center}
\bf Abstract
\end{center}
The Casimir energy of a massless scalar field is semiclassically
given by contributions due to classical periodic rays. The
required subtractions in the spectral density are determined
explicitly. The so defined semiclassical Casimir energy coincides
with that obtained using zeta function regularization in the cases
studied. Poles in the analytic continuation of zeta function
regularization are related to non-universal subtractions in the
spectral density. The sign of the Casimir energy of a scalar field
on a smooth manifold is estimated by the sign of the contribution
due to the shortest periodic rays only. Demanding continuity of
the Casimir energy under small deformations of the manifold, the
method is extended to integrable systems. The Casimir energy of a
massless scalar field on a manifold with boundaries includes
contributions due to periodic rays that lie entirely within the
boundaries. These contributions in general depend on the boundary
conditions. Although the Casimir energy due to a massless scalar
field may be sensitive to the physical dimensions of manifolds
with boundary, its sign can in favorable cases be inferred without
explicit calculation of the Casimir energy.

\bigskip


\newpage

\section{Introduction}

Classically, the energy of a field is always positive. That need
not be so for a Casimir energy ${\cal E}$, generally thought to
have its origin in the vacuum fluctuations of the
field\cite{Ca48}. The possibility of ${\cal E}$ being negative can
be understood from the fact that it is the \emph{difference} in
the infinite zero-point energies of the field for two systems.
Most calculations in the past were performed by directly
evaluating this difference of infinite zero-point energies. There
are many articles and texts that consider Casimir effects in this
manner. See for instance\cite{Mi93,MT97,Mi01} for an overview.

The ultra-violet divergence of the zero-point energy in general
reflects local properties of the system. From a path-integral
point of view the divergence is due to contributions from
arbitrarily short paths that begin and end at the same point.
These zero-length paths probe the local radius of curvature or,
for paths touching boundaries, local properties of the
boundary\cite{CD79, Fu89, Ja02, Mi02, Fu03}. In a few favorable
situations, ultra-short paths do not contribute to the
\emph{difference} of zero-point energies. This in particular is
the case for rigid disjoint boundaries that are moved relative to
each other\cite{Gi03}. One sometimes also considers idealized
boundaries whose local deformation does not cost energy. An
example of the latter are smooth, perfectly conducting metallic
surfaces of vanishing thickness\cite{Mi78} in three dimensions.
The (local) surface tension of such an ideal surface
vanishes\cite{BGH03,BD04} and no energy is required to deform it
locally.

A \emph{measureable} Casimir energy should not be extremely
sensitive to the (sometimes implicit) ultra-violet cutoff and
should depend on global characteristics of the system only. It
otherwise is difficult to disentangle the energy required to
change the system as a whole from purely local effects, for
instance due to changes in the local curvature of the space or in
the transmission of a boundary.

There thus either is no contribution from ultra-short paths to a
measurable Casimir effect or it must be possible to unambiguously
isolate these local contributions to the vacuum energy. Most
calculations of the Casimir energy based on spectral properties
and Green function methods do not separate length scales
explicitly. The regularization and subsequent subtraction of
divergent contributions often are difficult to motivate physically
and it is not always apparent how Casimir energies of different
systems can to be compared.

The semiclassical evaluation of Casimir energies advocated
in\cite{MS98} relies on an \emph{ab initio} separation of scales.
The approach separates the semiclassical contribution to the
Casimir energy due to quadratic fluctuations about
\emph{classical} periodic rays (paths) from all others. Classical
periodic rays are of (finite) extremal length and give a
semiclassical approximation to the Casimir energy that depends on
global characteristics of the system only. This part of the vacuum
energy is naturally finite and does not include ultra-violet
contributions from length scales that are much smaller than the
shortest classical periodic ray. This is one of the principal
conceptual differences to the "optical" approximation to Casimir
energies\cite{Ja04}. The latter (in principle) takes all {\it
closed} classical paths (not just periodic ones) into
account\footnote{Although this appears to be an improvement over
the semiclassical treatment, the optical approximation to Casimir
energies also only includes quadratic fluctuations about classical
rays. The optical approach in principle could provide a more
uniform approximation in some cases (but not in all) but
inherently is no more accurate than the semiclassical approach. An
objective comparison of the two methods\cite{SS05} is rather
difficult due to the numerical limitations and approximations of
this approach.}. Closed paths can be ultra-short in the vicinity
of surfaces and lead to divergent Casimir self-energies.  The
optical approximation therefore has mainly been used to obtain a
numerical estimate of the interaction energy for rigid bodies.

It was argued in\cite{MS98} that a semiclassical evaluation of the
Casimir energy is often particularly simple and gives the leading
asymptotic behavior when the Casimir energy is large. This is the
experimentally most accessible region of parameter
space\cite{Mo99}.

However, the desired separation of length scales may not always be
possible: changing the radius of a spherical shell invariably
changes the local curvature as well. The energy required to
achieve a change in radius in this case will include a possibly
divergent contribution from the local change in curvature.

This suggests dividing systems into classes: the \emph{difference}
in vacuum energy of any two systems within the same class being
finite. It would require an infinite amount of energy to compare
systems belonging to different classes. Within a particular class,
the finite Casimir energy has the universal interpretation of a
vacuum energy: differences in Casimir energy are the finite
differences in vacuum energy.

The spectral density $\rho(E;\dots)$ is assumed to be a
well-defined quantity for any system (at least for "free",
non-interacting fields). The ellipsis here stand for the space
${\cal M}$, the types of field, the boundary conditions that are
satisfied and any other qualifiers of the system. For systems
${\cal A}$ and ${\cal B}$ of the same class, the difference of
spectral densities,
\bel{diffrho}
\rho(E;{\cal A}-{\cal B}):=\rho(E;{\cal A})-\rho(E;{\cal B})\ ,
\fin
by definition has a finite first moment,
\bel{EAB}
-\infty< {\cal E}_{{\cal A}-{\cal B}}:=\frac{1}{2}\int_0^\infty
\rho(E;{\cal A}-{\cal B}) E dE=E_{{\rm vac}}({\cal A})-E_{{\rm
vac}}({\cal B})<\infty\ .
\fin
${\cal E}_{{\cal A}-{\cal B}}$ could be called the Casimir energy
of system ${\cal A}$ with respect to system ${\cal B}$. There
evidently are many equivalent definitions of the Casimir energy of
a system within a particular class -- they are distinguished by
the spectral density used as reference. The Casimir energy
determined by two such subtraction schemes, differs only by a
finite amount that is the same for any system of a class. Such
subtraction schemes are equivalent in all physical respects.

The semiclassical Casimir energy (SCE) is defined by a particular
subtraction $\rho_0(E)$ in each class.  I will take advantage of
the fact that the semiclassical spectral density $\rho(E)$ is the
sum of a part $\tilde\rho(E)$ determined by contributions from
periodic rays and a (often classical) remainder\cite{Gu90}
$\rho_0(E)$,
\bel{rho}
\tilde\rho(E;\dots)=\rho(E;\dots)-\rho_0(E;{\rm
class})=-\frac{1}{\pi}\lim_{\eps\rightarrow 0^+} {\rm Im}\, \tilde
g(E+i\eps;\dots)\ .
\fin
Here $\tilde g(E)$ is the part of the response function due to
classical periodic rays. For a scalar field the remainder
$\rho_0(E)$ at least includes the Weyl contribution to the
spectral density proportional to the volume of ${\cal M}$. In
addition $\rho_0(E)$ may depend on the type of field, the
curvature, boundaries as well as other characteristics\cite{CD79,
Fu89,BD04}.

The SCE ${\cal E}_c$ then is defined as,
\bel{Cascdef}
{\cal E}_c({\cal M})=\int_0^\infty \frac{E}{2}\tilde\rho(E) dE =
        -\frac{1}{\pi}\lim_{\eps\rightarrow 0}\int_0^\infty\frac{E}{2}{\rm Im} \tilde g(E+i\eps) dE\ .
\fin
Since the length of a periodic ray is finite, this contribution to
the vacuum energy is free of ultra-violet divergences and in
general is \emph{finite}\cite{MS98}. The SCE of\equ{Cascdef}  may
be taken to (at least approximately) represent the vacuum energy
within a class of systems for which the subtracted spectral
density $\rho_0(E)$ is the \emph{same}. It may happen (see the
example of the Laplace-Beltrami operator on a half-sphere of
Appendix~A) that a particular class has just one member. The
subtraction $\rho_0(E)$ in this case is not universal to several
($\geq 2$) systems. The finite Casimir energy one extracts in this
case is peculiar to a particular system and is physically quite
irrelevant: any change in the system requires infinite energy. In
several examples studied in Appendix~A such non-universal
subtractions are associated with poles in zeta function
regularization. The SCE of\equ{Cascdef} on the other hand
coincided with the Casimir energy of zeta function regularization
in all systems I studied for which the subtraction has a more
universal meaning.

Although\equ{Cascdef} does not directly refer to $\rho_0(E)$, this
implicit subtraction in the spectral density determines the class
of systems and thus, in effect, the usefulness of the SCE. Other
approaches, such as zeta function regularization often give finite
answers without specifying what has been subtracted. Still other
approaches, such as heat kernel expansion, subtract terms whose
physical implications are not entirely clear\cite{Fu89, Fu03} and
the question whether one gains or looses vacuum energy by
transforming an elongated ellipsoid into a sphere is difficult to
answer. To escape this conundrum in the interpretation of a
Casimir energy, Power\cite{Po64} long ago considered a large
rectangular box with a moveable wall to define the original
Casimir energy\cite{Ca48} for two parallel conducting plates
unambiguously. He in effect was considering a class of systems
that all have the same total volume, total surface area, edge
length and number of corners. We will see in
Section~\ref{parallele} that the implicitly subtracted spectral
density $\rho_0(E)$ for a three-dimensional parallelepiped in fact
only depends on these characteristics. $\tilde\rho(E)$ of a
parallelepiped can again be expressed in terms of periodic
orbits\cite{BB70,Lu71,AW83} only.

Restricting the validity of a Casimir energy to a certain class of
spaces for which the same subtraction in the spectral density
gives a finite Casimir energy in this sense generalizes Power's
procedure to slightly less obvious situations. As the example in
Appendix~A of a massless scalar field on $S_4$ demonstrates,
(universal) subtractions can go beyond Weyl terms and for instance
include contributions proportional to the integral of the (local)
curvature over the whole space.

As emphasized in a perturbative setting by Barton\cite{Ba01}, the
physical interpretation of a Casimir energy depends almost
entirely on the (implicit) subtraction. This is readily
illustrated by a spherical cavity in three dimensions. The
significance\cite{Ca53} of the electromagnetic Casimir energy
(which was found to decrease with the radius of the
cavity\cite{Mi78}), relies on the fact that this Casimir energy
actually determines the physical pressure on the spherical surface
of the cavity. This conclusion is possible only if the (implicit)
subtractions in $\rho_0(E)$ do not depend on the surface area of
the boundary. The finite Casimir energy otherwise could only be
used to obtain the vacuum energy difference between cavities of
the same surface area and \emph{would not} determine the pressure
on the cavity surface. That a subtraction proportional to the
surface area is not required in the electromagnetic case, is due
to the ideal metallic boundary conditions\cite{BGH03,BD04}. The
situation is less favorable for a \emph{scalar} field\cite{Mi02}
satisfying Dirichlet boundary conditions on such a spherical
surface. The non-universality of the required subtraction was
emphasized in\cite{Ja02}.

Defining the Casimir energy in terms of contributions due to
periodic orbits rather than by any other subtraction of the
spectral density has the advantage that this finite part of the
vacuum energy may often be evaluated \emph{approximately}. This is
of practical use in situations where the exact spectrum is not, or
is only numerically, known. A rather crude approximation will give
an estimate of the sign of the SCE in\equ{Cascdef} without
detailed knowledge of the periodic rays themselves.

The sign of Casimir energies is one of its many puzzles. Without
explicit calculation, determining the sign of the
\emph{difference} of two divergent vacuum energies in general is
quite hopeless. Obtaining the sign of the SCE on the other hand is
much more promising due to the geometrical nature of this
definition. The overall phase of the contribution to the response
function from a particular periodic ray is given by a topological
winding number\cite{CRL90}. I will argue that the sign of the SCE
can often already be inferred from the shortest periodic rays that
contribute.

I first illustrate the approach for single valued (bosonic) fields
on smooth $d$-dimensional manifolds without boundary such as $S_d$
and $T_d$.  I then generalize to manifolds with boundaries on
which the bosonic field satisfies Dirichlet or Neumann conditions.
Several examples show that classical periodic rays within the
boundary must also be considered. In general the contribution of
these rays to the SCE depends on the boundary condition. When the
boundary is not smooth, as for a parallelepiped, contributions due
to periodic rays in even lower dimensional spaces have to be
included as well.

\section{General Spaces without Boundary} The conceptually simplest Casimir
energy probably is that due to a massless single valued bosonic
field on a smooth $d+1$-dimensional Riemannian space-time without
boundary. I will assume that the metric is static in a particular
frame, i.e. that it makes sense to speak of a $d$-dimensional
spatial manifold ${\cal M}$ and of the energy of a particle.
Periodic rays follow geodesics on ${\cal M}$ that close on
themselves. The classical action for a periodic ray $\gamma$ then
is,
\bel{action}
S_\gamma=\oint {\bf p}\cdot{\bf dx}= p(E) L_\gamma,
\fin
where $\tau_\gamma=\partial S_\gamma/\partial E$ is the time for
the ray to return to its starting point on ${\cal M}$. For a
massless particle moving at the speed of light, $\tau_\gamma=
L_\gamma/c$ and thus $p(E)=E/c$.  Note that for periodic rays
$\tau_\gamma$ is an integer multiple of the primitive period
$t_\gamma$.

The contribution $\tilde g(E)$ of \emph{isolated} periodic rays to
the response function is of the form\cite{Gu90} ,
\bel{response}
\tilde g(E)=\frac{1}{i\hbar}\sum_\gamma A_\gamma t_\gamma e^{i
E\tau_\gamma/\hbar-i\sigma_\gamma\pi/2}\ .
\fin
In\equ{response} the amplitude $A_\gamma$ is determined by the
monodromy matrix associated with the ray $\gamma$. It is a
geometric quantity that (for massless particles in vacuum) does
not depend on their energy $E$. $A_\gamma$ furthermore is positive
and real by definition. The integer $\sigma_\gamma\geq 0$ is the
Maslov-like index of the stable and unstable manifolds of the
periodic ray\cite{CRL90}. Important for us is that this index is a
topological winding number. As such it is an additive integer that
scales directly with the number of times an orbit is iterated. For
practical calculations it will be useful that $\sigma_\gamma$ may
be written as the sum,
\bel{sigma}
\sigma_\gamma=\mu_\gamma+\nu_\gamma\ ,
\fin
of the number of conjugate points $\mu_\gamma$ between the initial
point ${\bf x}$ and the final point ${\bf x}'={\bf x}$ of the
periodic ray and of an integer $\nu_\gamma$ associated with the
stability of the periodic orbit.

$\mu_\gamma$ gives the total phase retardation $\mu_\gamma\pi/2$
due to conjugate points (for manifolds without boundaries)
encountered by the periodic ray.
 $\nu_\gamma$ in\equ{sigma} is the number of
\emph{negative} eigenvalues of the matrix ${\bf W}$ of second
variations of $L_\gamma$ with respect to a change of the initial
(=final) point of the periodic ray\cite{Gu90},
\bel{W}
\left.\delta L_\gamma({\bf x}+\delta {\bf y},{\bf x}'+\delta {\bf
y})\right|_{{\bf x}={\bf x}'}=\delta {\bf y}^T\cdot {\bf W}({\bf
x})\cdot \delta {\bf y}\ .
\fin
Since one of the eigenvalues of the $d\times d$ matrix ${\bf W}$
always vanishes, $0\leq \nu_\gamma\leq d-1$.

If all periodic rays are isolated, one can insert\equ{response} in
the definition of\equ{Cascdef}. Upon performing the energy
integral, the SCE due to only isolated periodic rays is of the
form,
\bel{CasimirE}
{\cal E}_c({\cal M})=-\hbar\sum_\gamma \cos(\sigma_\gamma\pi/2)\frac{A_\gamma t_\gamma}{2\pi\tau_\gamma^2}\\
\fin
The sign of the contribution of a particular periodic ray to the
SCE is determined by the integer $\sigma_\gamma$. Remarkably,
periodic rays with odd $\sigma_\gamma$ do not contribute to the
Casimir energy. Although an expression like \equ{CasimirE} is
valid only for \emph{isolated} periodic rays, it can also be used
to estimate the \emph{sign} of the Casimir energy of integrable
systems. The expression of Berry and Tabor\cite{BT76} for the
spectral density of an integrable system in terms of periodic rays
is more appropriate (see below), but integrable systems are
singular in the sense that small deformations of the manifold
${\cal M}$ destroy the symmetries and generically result in
isolated periodic rays. The expression of\equ{CasimirE} is robust
in the sense that the SCE changes continuously and in particular
generally does not change sign if the deformation is small enough.
Slightly deforming ${\cal M}$ to isolate the orbits therefore
should allow us to obtain the sign of ${\cal E}_c$
from\equ{CasimirE} even for integrable systems. In support of this
conjecture note that the integer $\sigma_\gamma$ of an individual
periodic ray in\equ{response} is a winding number that changes
only when a new caustic appears or the stability of the periodic
ray changes. If the contribution of a periodic ray to the Casimir
energy does not vanish, its sign should not change for
sufficiently small deformations of the manifold.

A unique determination of the sign of ${\cal E}_c$ is possible
when $\cos(\sigma_\gamma\pi/2)$ does not depend on the periodic
ray $\gamma$. In less favorable situations I resort to finding the
sign of the contribution due to the \emph{shortest} periodic rays
to\equ{CasimirE}.  The contribution in\equ{CasimirE} of a periodic
ray that winds $n$ times about the geodesic generally decreases in
magnitude as $1/n^2$ and in some cases decreases even faster. If
the contribution from primitive periodic rays dominates the SCE
in\equ{CasimirE}, I will estimate its sign by that of the
\emph{shortest} primitive periodic rays. For some spaces (see for
instance $S_{2n}$ below), short primitive rays do not contribute
to the Casimir energy at all or may give contributions of either
sign. The overall sign of the Casimir energy in this case is
ambiguous and this estimate fails. One nevertheless might expect
${\cal E}_c$ to be rather small in magnitude in such situations
and I will write ${\cal E}_c\sim 0$ when the sign cannot be
determined from the shortest periodic rays.

\subsection{$d$-dimensional Tori and Spheres}\label{torispheres}
Obtaining the sign of the SCE is straightforward for a massless
scalar on a $d$-dimensional torus $T_d=S_1\times
S_1\times\dots\times S_1$. The curvature of $T_d$ vanishes and it
is a space without boundary. The subtracted spectral density
therefore is the Weyl-term proportional to the volume of $T_d$
only. However, due to the translational symmetries this is an
integrable system and classical periodic rays are not isolated. To
estimate the sign of the SCE of a torus using\equ{CasimirE}, one
has to deform it slightly. Such a deformation generally destroys
all symmetries and gives isolated periodic rays, the shortest of
which resemble periodic rays of the original torus on its shortest
cycles. If the curvature remains sufficiently small on the
deformed torus, the number of conjugate points along a primitive
orbit continues to vanish. The length of the shortest periodic
rays furthermore is a minimum by definition. One thus obtains
$\mu_\gamma=\nu_\gamma=\sigma_\gamma=0$ for the shortest periodic
rays of a slightly deformed torus. They all give a negative
contribution to the Casimir energy in\equ{CasimirE}. Since this
sign does not depend on the particular deformation of the torus,
one can be confident that,
\bel{sTd}
{\cal E}_c(T_d)<0\ {\rm for~all}\ d=1,2,\dots
\fin
This  sign is in agreement with that of the Casimir energy due to
a massless scalar field satisfying \emph{periodic} boundary
conditions on the hyper-surface of any $d$-dimensional
parallelepiped obtained by explicit calculation\cite{AW83}.

\equ{CasimirE} indicates that the SCE may in principle be of
either sign for manifolds without boundaries. A non-trivial
example is the Casimir energy of a massless scalar field confined
to a spherical shell $S_d$ of dimension $d$ and radius $R$.
Periodic rays follow great circles of radius $R$ on $S_d$. They
again are not isolated and the $d-1$ dimensional cross section of
a bundle of initially parallel geodesics is reduced to a point at
two anti-podes. Any starting point on a geodesic of $S_d$ also is
a conjugate point of order $d-1$, i.e. it is
self-conjugate\cite{Bo88}. One can avoid the associated
complications by slightly deforming $S_d$ in a generic fashion.
The primitive periodic rays of the deformed sphere should still
resemble original geodesics on $S_d$, but the starting point
generally no longer is self-conjugate. A pencil of rays emanating
from a point on a geodesic also generally no longer meets in a
single focal point. For a sufficiently  small deformation, focal
points of the sphere will have been resolved into a series of
$d-1$ closely spaced conjugate points of first order. On the
\emph{shortest} primitive periodic ray, there generically are only
$d-1$ conjugate points, the next bunch of $d-1$ conjugate points
occurring just \emph{after} completion of a full revolution. The
reason is that the curvature along the shortest primitive ray in
general is slightly too \emph{small} to lead to more than one
intersection\footnote{The geodesic distance $\eta(s)$ between two
nearby geodesics satisfies the linear second order equation
$d^2\eta(s)/ds^2=-\kappa(s)\eta(s)$, where $s$ is the arc length
along the ray and $\kappa(s)$ is the Gaussian curvature at $s$. If
$\kappa(s)<4\pi^2/L_\gamma^2$ two geodesics at most meet once.
This condition generally holds for the \emph{shortest} primitive
ray of a slightly deformed sphere.} -- as for a periodic ray about
the waist of a slightly elongated ellipsoid. The \emph{shortest}
periodic rays furthermore are of minimal length and therefore are
stable and $\nu_\gamma=0$ in this case. The total phase
retardation of the shortest primitive rays thus is $\sigma_\gamma
\pi/2=\mu_\gamma\pi/2=(d-1)\pi/2$. Their contribution to the
Casimir energy of a (deformed) sphere in\equ{CasimirE} vanishes in
even dimensions and is of alternating sign in odd dimensions.
Assuming that the contribution from the shortest periodic rays
dominate in\equ{CasimirE}, one thus estimates that,
\bel{signSd}
{\cal E}_c(S_d)\propto -\cos(\pi(d-1)/2) \left\{\begin{array}{ll}    <0,& {\rm for}\ d=1\mod{4}\\
                      \sim 0,& {\rm for\ even}\ d\\
                      >0,& {\rm for}\ d=3\mod{4}\\
                      \end{array}\right.
\fin
The explicit calculations for $d\leq 4$ of Appendix~A confirm this
not very intuitive pattern for the sign of the Casimir energy of a
massless scalar field on low-dimensional spheres. A little
surprisingly, the Casimir energy vanishes \emph{exactly} for
$S_2$, $S_4$ and in fact any $S_{{\rm even}}$. Since our somewhat
crude estimation only takes contributions from the shortest
primitive rays into account, it cannot in itself predict a
vanishing Casimir energy. When the contribution from the shortest
periodic rays vanishes, determining the overall sign of the
Casimir energy becomes much more involved. To conclude that the
Casimir energy vanishes one has to show that it is positive for
some (small) deformations of the manifold and negative for others.
Although this can be shown for $S_{{\rm even}}$, the argument is
no longer "simple" and I will not pursue it any further.

It is amusing to consider further stretching the sphere to an
elongated cigar-like shape and eventually a long cylinder with end
caps. The subtractions have to remain constant during this
deformation. In two and three dimensions, this amounts to keeping
the total surface area, respectively volume, constant during the
deformation. In higher dimensions, certain moments of the
curvature and other global characteristics also must not change.
As the curvature decreases, the $d-1$ conjugate points on the
shortest (and thus stable) periodic rays move beyond the end of
the primitive orbit when the sphere is stretched sufficiently.
$\sigma_\gamma\rightarrow 0$ for these orbits and they eventually
dominate and lead to a negative Casimir energy for a (slightly
deformed) long cylindrical $d$-dimensional surface (for
essentially the same reason as the short rays on a torus). Since
the length of the shortest periodic rays on a very elongated
cylinder is much less than on a sphere of the same $d$-dimensional
volume, the Casimir energy furthermore increases in magnitude
during this deformation\footnote{It would be erroneous to compare
the Casimir energy densities of a cylindrical surface and a
spherical one of the same radius when the subtractions are not the
same.}. A free massless scalar field thus would collapse $S_d$ to
a filament.

\section{Integrable Systems}\label{integrablesys}

A quantitative comparison with the Casimir energy obtained by
other methods generally is possible only for integrable systems,
for which the true spectrum is explicitly or
implicitly\cite{KNS68} known.

In the examples studied in Appendix~A the SCE coincides with that
obtained by zeta function regularization when the analytic
continuation is uniquely defined. When poles arise and zeta
function regularization gives ambiguous answers, the associated
subtraction in the spectral density is not universal either.
Universal subtractions on the other hand may include terms in
$\rho_0(E)$ that depend on global characteristics of the curvature
[as for the Casimir energy on $S_4$ in Appendix~A].

All our estimates of the sign of the SCE rely on the dispersion
relation $p(E)=E/c$ of a \emph{massless} particle, since the
integral in\equ{Cascdef} had to be performed explicitly  to arrive
at\equ{CasimirE}. The sign of the SCE in general will differ for
other dispersion relations, as for instance for the spectrum of
the Laplace-Beltrami operator without curvature correction. As
argued in Appendix~A the latter is not the spectrum of a massless
scalar\cite{Sc81,Gu90}. Using zeta function regularization, the
Casimir energy in this case is negative for all $S_d, d\leq
4$\cite{El94}, but the Casimir energy of $S_3$ is ambiguous due to
a pole contribution.

The explicit calculations of Appendix~A strongly suggest that the
SCE of a massless scalar on $S_d$ vanishes \emph{exactly} for even
dimension $d$: the integrands in this case are polynomials in
$(ER)^2$ only. The subtraction $\rho_0(E)$ for $S_{{\rm even}}$
otherwise would not be universal. The integral over the energy of
the response function (after Wick rotation) has vanishing
imaginary part and there is no contribution to the Casimir energy
from {\it any} periodic ray on $S_{{\rm even}}$.

Although this supports the previous estimate of the sign of the
SCE for low-dimensional spheres, \equ{CasimirE} requires periodic
rays that are isolated and thus cannot be directly applied to
integrable systems. Let us therefore obtain an expression for the
SCE that can be used to determine its sign in the (rather special)
case of an integrable system.

Action-angle variables are the canonical phase-space coordinates
of integrable systems. The SCE for a $d$-dimensional integrable
system is found by applying Poisson's formula\cite{Gu90,BT76},
\bel{Casint}
{\cal E}_c\left({{\rm integrable}\atop{\rm system}}\right)=
\frac{1}{2\hbar^d}\sum_{\bf m\neq 0}\int H({\bf I})\; e^{2\pi i
{\bf m}\cdot[{\bf I}/\hbar-\bsbeta/4]} {\bf dI}\ .
\fin
Here $H({\bf I})$ is the classical hamiltonian expressed in terms
of the actions ${\bf I}=(I_1,I_2,\dots,I_d)$ and ${\bf m}$ is a
$d$-dimensional vector of integers. The summation in\equ{Casint}
extends over all such vectors except ${\bf m}={\bf 0}=
(0,0,\dots,0)$. This contribution has been subtracted by the
"classical" spectral density,
\bel{rhoclass}
\rho_0(E)=\int \delta(E-H({\bf I})) \frac{{\bf dI}}{\hbar^d} \ .
\fin
\equ{Casint} expresses the SCE of an integrable system as a sum
over classical periodic trajectories on the invariant torii. The
classical action of a trajectory with winding numbers ${\bf m}$
about each of the cycles of the invariant torus is $S({\bf
m})=2\pi {\bf m}\cdot{\bf I}$. The correction to the classical
action proportional to $\hbar$ is linear in the winding numbers
${\bf m}$. It is a topological quantity that determines how
periodic orbits on the invariant torus are projected onto physical
coordinate space\cite{Ke58}. The vector $\bbeta$ of
Keller-Maslov\cite{Ma72} indices gives the phase loss from
caustics on a periodic orbit (see below). Note that $\bbeta$ is a
geometrical quantity that does not depend on ${\bf I}$.

The integrals over the actions of\equ{Casint} are evaluated
semiclassically at stationary points $\bar{\bf I}(E,{\bf m})$ of
the classical action on the energy surface -- the vector ${\bf m}$
is normal to the energy surface $H({\bf I})=E$ at ${\bf
I}=\bar{\bf I}(E,{\bf m})$. For given energy $E$, the $d-1$
integrations along the (compact) energy surface are performed
semiclassically by choosing a local frame of actions at ${\bar{\bf
I}}$ for which one axis, say that of $I_1$, is in the direction of
$\bf m$ and all others are tangent to the energy surface.  Care
must be taken with zero modes of the matrix of second derivatives
\bel{hess}
H_{ij}=\partial^2 H/\partial \bar I_i\partial \bar I_j,\
i,j=2,\dots,d\ .
\fin
The integral over the $\nu_0$ dimensional subspace of the zero
modes of $H_{ij}$ is stationary only when the corresponding
frequencies vanish. The sum over ${\bf m}$ is thereby reduced to
one over the $(d-\nu_0)$-dimensional vectors ${\bf n}$ that are
orthogonal to the zero modes. Denoting the $\nu_0$-dimensional
volume of this classical moduli-space by $V_{\nu_0}(E,{\bf n})$,
the semiclassical result for the $d-1$ integrations along the
energy surface in\equ{Casint} is,
\bel{Casintf}
{\cal E}_c\left({{\rm integrable}\atop{\rm system}}\right)=
\sum_{{\bf n}\neq {\bf 0}}\int_0^\infty \frac{E dE}{2
\omega\hbar^d} \frac{V_{\nu_0}(E,{\bf n}) \
e^{i\pi(\nu_--\nu_+)/4}}{\sqrt{|{\bf
n}/\hbar\omega|^{d-1-\nu_0}\,|\det^\prime H_{ij}|} } \ e^{2\pi i
{\bf n}\cdot[{\bf \bar I}/\hbar - \bsbeta/4]}\ .\\
\fin
The primed determinant here means that the $\nu_0$ zero-modes have
been omitted in its calculation. $\nu_+$ and $\nu_-$ are the
number of positive and negative eigenvalues of $H_{ij}$,
$\nu_++\nu_-+\nu_0=d-1$. The frequency $\omega=|{\bf
\omega}|=|\nabla_I H|_{I=\bar I(E,{\bf m})}$ for a \emph{massless}
particle in fact does not depend on its energy.

The energy dependence of the integrand in\equ{Casintf} is made
explicit by noting that the classical action of a massless field
is $S(E,{\bf n})=|2\pi {\bf n}\cdot {\bf \bar I}|=E L_{{\bf n}}/c$
where $L_{{\bf n}}>0$ is the length of the periodic ray. For
dimensional reasons, $\det^\prime H_{ij}(E)$ scales as
$(E/\omega^2)^{1-d+\nu_0}$ and $V_{\nu_0}(E)$ scales as
$(E/\omega)^{\nu_0}$. The integration over the energy $E$
in\equ{Casintf} then gives the SCE of an integrable system without
boundaries in the form,
\bel{Casintfin}
{\cal E}_c\left({{\rm integrable}\atop{\rm system}}\right)= -\hbar
c \sum_{{\bf n}\neq {\bf 0}}\left(\frac{c}{L_{\bf n}}\right)^{d+1}
A_{\bf \hat n}\ \cos{(\pi(\beta-\nu_0-\nu_-)_{\bf n}/2)} \ .
\fin
To perform the integral in\equ{Casintf}, the integer
\bel{identify}
\tilde\sigma_{{\bf n}}=(\beta-\nu_0-\nu_-)_{{\bf n}}\ ,
\fin
must not depend on $E$. Since geodesics do not depend on the
energy of a ray, the number of non-positive eigenvalues of
$H_{ij}$ does not depend on $E$ for massless particles. The
Maslov-Keller index $\beta_{{\bf n}}={\bf n}\cdot\bbeta$ depends
only on ${\bf n}$ by construction. Note that any \emph{constant}
angle variable of a closed geodesic implies that the hamiltonian
does not depend on the conjugate action. The dimension
$0\leq\nu_0\leq d-1$ of the space of zero modes of $H_{ij}$ can
thus often be obtained by inspection.

The amplitude $A_{\bf \hat n}$ in\equ{Casintfin} is positive by
construction and I have absorbed all dependence on the scale
$|{\bf n}|$ in the length $L_{{\bf n}}$ of a class of periodic
rays. $A_{\bf \hat n}$ thus is a function of the dimensions of the
integrable system (such as the volume of the space) that do not
scale with the length of the periodic orbit. The leading
contribution to the Casimir energy of an integrable system
in\equ{Casintfin} generally is from the shortest rays that
contribute to\equ{Casintfin}. The sign of this contribution
depends on the Keller-Maslov index $\beta_{{\bf n}}$ of the rays
as well as the number of non-positive eigenvalues of $H_{ij}$. The
sign of the contribution from a particular class ${{\bf n}}$ of
periodic rays of an integrable system is again given by,
\bel{signint}
-\cos(\tilde\sigma_{{\bf n}}\pi/2)\ ,\nonumber\\
\fin
with $\tilde\sigma_{{\bf n}}$ defined by\equ{identify}. The SCE
of\equ{Casintfin} for an integrable system thus is of a remarkably
similar form as the one in\equ{CasimirE} for a system with only
isolated periodic rays. It is tempting to identify
$\tilde\sigma_{{\bf n}}$ in\equ{identify} with $\sigma_\gamma$
of\equ{sigma}. However, while both $\sigma_\gamma$ and
$\tilde\sigma_{{\bf n}}$ are topological quantities,
$\sigma_\gamma$ in Gutzwiller's expression for the contribution to
the response function of isolated rays is a topological property
of the ray\cite{CRL90}, whereas $\tilde\sigma_{{\bf n}}$
in\equ{identify} is a property of a whole continuous family ${\bf
n}$ of periodic rays. The two integers depend differently on the
winding number of the periodic rays. One nevertheless can argue
that they coincide for the \emph{shortest} (and thus primitive)
periodic rays.

The Keller-Maslov index $\beta_{{\bf n}}$ of an integrable system
is given by the number of caustics a periodic ray encounters (for
manifolds without boundary). These caustics are created by the
family of rays it is a member of. $\beta_{{\bf n}}$ thus is a
topological property of the family of rays that generally does
\emph{not} equal $\mu_\gamma$, the number of conjugate points on a
single ray of that family. For one, $\beta_{{\bf n}}$ does not
depend on the starting point of the ray and is a well-defined
integer even when this point happens to be self-conjugate (as in
the case of spheres). $\beta_{{\bf n}}$ is proportional to the
number of times the periodic rays wind about the closed geodesic.
The number of positive eigenvalues of $H_{ij}$ on the other hand
is a statement about the curvature at a particular point on the
energy surface determined by the direction of its normal ${\bf
\hat n}$. Neither this point nor the energy surface change with
the magnitude of ${\bf n}$. $(\nu_0+\nu_-)=(d-1-\nu_+)$ thus
\emph{does not} depend on the winding number of the periodic ray.
For $\sigma_\gamma$ to coincide with $\tilde\sigma_{{\bf n}}$ for
\emph{all} periodic rays $\gamma$ as the integrable system is
slightly deformed, the number of non-positive eigenvalues of
$H_{ij}$ has to vanish (as for a torus). However, the sign of the
Casimir energy in general does not change under small
deformations, if the phase of the \emph{shortest} periodic rays
remains the same. This observation gives the desired connection
between $\sigma_\gamma$ and $\tilde\sigma_{{\bf n}}$: the two have
to coincide for the \emph{shortest} periodic rays of an integrable
system and its (sufficiently small) deformation. Below I show that
this  is the case for the previous example of spheres and tori.

\subsection{Covering Spaces}\label{coveringspaces} The previous
arguments imply that the sign of the Casimir energy can often be
inferred from the Keller-Maslov index $\beta_\gamma$ of periodic
rays of integrable systems. [$\gamma$ here is some representative
of the class ${{\bf n}}$.] It thus is important to have a reliable
and transparent determination of this index. Keller\cite{Ke58}
gives a geometrical construction that generalizes to manifolds
with boundaries. I recall points of the construction that are
relevant here.

A solution $S(\vec q,t)$ of the Hamilton-Jacobi equation $H(\vec
q, \vec p=\grad S,t)=E$ for constant energy $E$ generally is
multiply valued and so may be the momentum $\vec p=\grad S$
itself. At any point $\vec x$ the momentum $\vec p=\grad S$ of a
solution only has a finite number of branches, say $m$ of them.
One constructs an $m$-sheeted covering space on which $\vec
p=\grad S$ is a single-valued function by associating each of the
branches of $\grad S$ with a separate sheet. Any two different
sheets $i$ and $j$ are joined together on sub-manifolds on which
the momenta coincide $\left.\vec p\right|_i=\left.\vec
p\right|_j$. Such a sub-manifold generically is a caustic of the
rays or a boundary in the original space. [If $\grad S$ is defined
on only part of $x$-space then only this part is covered.] The
advantage of Keller's construction is that $\vec p=\grad S$
becomes a single-valued function on the covering space. A family
${\bf n}$ of periodic rays does not intersect on the covering
space.

The semi-classical Casimir energy of the integrable system is
obtained by considering the periodic rays on this covering space.
The phase is retarded by $m\pi/2$ whenever a ray crosses a caustic
of $m^{\rm th}$ order in passing from one sheet to another. The
positive integer $m$ is the number of dimensions by which the
cross-section of a tube of nearby trajectories is reduced at the
caustic. Keller's construction does not of itself provide the
order of a caustic. The latter must be inferred from the behavior
of a bundle of nearby rays.

An example is the construction of this covering space for geodesic
motion on a $d$-dimensional sphere $S_d$. The $SO(d+1)$ symmetry
of $S_d$ implies that geodesics lie in a hyperplane of the ${\bf
R}_{d+1}$ it is embedded in. The vector orthogonal to this
hyperplane makes a certain angle $\theta$ with the vertical which
is the inclination of the geodesic. It can be chosen as one of the
angles of the action-angle variables and geodesics of $S_d$ thus
fall into distinct classes characterized by their inclination. A
family of periodic rays of the same inclination covers an annulus
of $S_d$ that is bounded by two $S_{d-1}$ hyper-surfaces. These
two $d-1$-dimensional caustics are around the polar regions of
$S_d$. Every closed geodesic in the family of solutions with given
inclination touches the "upper" and "lower" caustic once. One
constructs the covering space by joining two sheets of this
annulus of $S_d$ at the two caustics. A periodic ray of the given
inclination passes from one sheet to the other every time it
crosses one of these caustics.  In this covering space a family of
rays of fixed inclination does not intersect. The number of times
a periodic ray of fixed inclination winds about the annulus also
is the number of times the periodic ray passes through {\it both}
caustics. These caustics are of order $d-1$, because the cross
section of a $d-1$-dimensional tube of geodesics vanishes when
they intersect a given periodic ray at the caustic. The phase loss
of a periodic ray that winds about the sphere $n$ times and
crosses $2 n$ caustics is $2 n(d-1)\pi/2$ and thus $\beta_\gamma=
2 n(d-1)$ for any family of periodic rays of fixed inclination.
The hamiltonian depends on the action conjugate to the angle
describing the motion on a great circle only (the magnitude of
angular momentum in the 2-dimensional case). Therefore $\nu_0=d-1$
and $\nu_-=0$ in this case leading to $\tilde\sigma=(2n-1)(d-1)$.
For even dimensions $d$, $\tilde\sigma$ thus is odd for all $n$
and the contribution to the SCE of \emph{all} periodic rays
according to\equ{signint} vanishes. This agrees with our previous
estimation of the signs in\equ{signSd} based on\equ{CasimirE} and
proves that the SCE of even-dimensional spheres indeed vanishes.
Appendix~A presents explicit results for $0<d\leq 4$.

Note that the covering space for rays on a  $d$-dimensional torus
is trivial since the momentum is single-valued. The hamiltonian in
this case furthermore depends on the modulus of the momentum in
each direction and $H_{ij}$ is positive definite. $\nu_0+\nu_-=0$
for a massless particle on the torus and $\tilde\sigma_{{\bf n}}
=0$ for any periodic ray. The Casimir energy of a torus therefore
is always negative, as was already found by deforming it and
using\equ{CasimirE}.

\section{Manifolds with Boundaries}

Estimates of the sign of the SCE are more difficult for manifolds
with a boundary on which the scalar field satisfies some
conditions. I here consider Dirichlet and Neumann boundary
conditions only. Since classical rays reflect specularly at a
boundary, the basic strategy is to glue copies of the original
manifold at the boundaries and consider the resulting covering
manifold without boundary. At a boundary $\vec p=\grad S$ is
discontinuous and there are (at least) two values for the momentum
at any point near a boundary. By constructing the covering
manifold for which $\vec p=\grad S$ is single-valued, one thus can
treat a boundary in much the same way as a caustic. The phase
retardation at a boundary depends on the boundary condition. The
phase loss is $\pi$ for Dirichlet and $0$ for Neumann boundary
conditions. This ensures the correct behavior of semiclassical
Green functions near the boundary.

However, an additional correction to the semi-classical Casimir
energy arises from periodic rays of the covering manifold that lie
(entirely) within the boundary. The sign and magnitude of this
additional contribution can be essential in determining the sign
of the semi-classical Casimir energy and its dependence on the
boundary condition.  Let us examine some simple examples.

\subsection{Semiclassical Casimir Energy of a $d$-dimensional Half-Sphere}
Consider first the semi-classical Casimir energy of a half-sphere
in $d$-dimensions. Space in this case is just $S_d$ cut in half at
the equatorial $S_{d-1}$. Classical periodic rays in this case lie
on two halves of great circles of the original $S_d$ with the
\emph{same} inclination that intersect on the equatorial $S_{d-1}$
of the sphere. The momentum at any point on the
half-sphere-annulus covered by a family of rays with fixed
inclination therefore can take up to 4 values and one needs a
4-sheeted covering of this space to make $\grad S$ single-valued.
This covering space is constructed in two steps. One first doubles
the half-sphere and joins the two sheets at the equators to form
an $S_d$. On this boundary-less double-covering of the
half-sphere, geodesics are again great-circles and one again
introduces two coverings for each annulus of $S_d$. Note that this
last operation \emph{doubles} the equatorial (boundary) $S_{d-1}$.
One of these "equators" is where the upper and lower parts of the
\emph{inner} annulus join, the other is where the corresponding
parts of the \emph{outer} annulus join. This doubling of the
boundary of the half-sphere cannot be avoided if the momentum on
the covering space is to be single-valued.

The periodic rays of this 4-sheeted covering of the
$d$-dimensional half-sphere evidently are those already found for
the 2-sheeted covering space of $S_d$. The only difference is that
the phase of a ray may be retarded by $\pi$ at every crossing of
an "equator" (for Dirichlet boundary conditions). Since a periodic
ray, however, crosses the "equators" an even number of times, one
is tempted to conclude that the semi-classical Casimir energy of
the half-sphere is just half the semi-classical Casimir energy of
$S_d$, irrespective of whether Neumann or Dirichlet boundary
conditions have been imposed.

However, this argument ignores classical periodic rays of the
covering space that lie entirely \emph{within} the boundary. The
contribution of such rays in general depends on the imposed
boundary condition. Since the field vanishes on a boundary with
Dirichlet's condition, classical periodic rays that lie entirely
within the boundary should not contribute to Green's function and
the spectral density. Their contribution to the spectral density
of the manifold without boundary has to be subtracted in this
case.

Often, as in the case of a sphere, symmetry arguments can be
invoked to relate the Casimir energy for Dirichlet and Neumann
boundary conditions. Due to symmetry under reflections about the
equatorial plane, eigenfunctions of the hamiltonian can be chosen
to satisfy either Neumann or Dirichlet boundary conditions at the
equator of $S_d$. The sum of the Casimir energies on the
half-sphere ${\cal E}(S_2/2;N)$ and ${\cal E}(S_2/2;D)$ for
Neumann, respectively Dirichlet boundary conditions, therefore is
the Casimir energy of the full sphere,
\bel{CasNpD}
{\cal E}_c(S_d/2;N)+ {\cal E}_c(S_d/2;D)={\cal E}_c(S_d) \ .
\fin
The difference ${\cal E}_c(S_d/2;N)-{\cal E}_c(S_d/2;D)$ is due to
contributions to the Casimir energy from the boundary, i.e. due to
periodic rays on the equator. The magnitude of this contribution
in general is difficult to obtain without explicit calculation. It
is \emph{not} simply related to the Casimir energy of a $d-1$
dimensional sphere, because ${\cal E}_c(S_{d-1})$ does not include
fluctuations transverse to the equator. One nevertheless can argue
the sign of the difference, that is whether Neumann or Dirichlet
boundary conditions lower the Casimir energy. The point is that
families of periodic rays in the vicinity of the equator of $S_d$
are similar to those on $S_{d-1}$ except that they nevertheless
pass two caustics of order $d-1$ -- rather than of order $d-2$ as
for $S_{d-1}$ -- in every revolution. This gives an additional
phase loss of $\pi$ for every revolution no matter how close to
the equatorial hyperplane the periodic rays are. Since the sign of
the contribution of a periodic ray to the Casimir energy of
$S_{d-1}$ is given by $\tilde\sigma(S_{d-1})=(2n-1)(d-2)$, the
additional phase loss of $n\pi$ on $S_d$ changes the index to
$\tilde\sigma({\rm equator}\ S_d)=(2n-1)(d-2)+2n$. From this one
obtains that,
\bel{CasNmD}
{\cal E}_c(S_d/2;N)-{\cal E}_c(S_d/2;D)\sim{\cal E}_c({\rm equator})\left\{\begin{array}{ll}<0,& {\rm for}\ d=0 \mod{4}\\
                      =0,& {\rm for\ odd}\ d\\
                      >0,& {\rm for}\ d=2 \mod{4} \\
                      \end{array}\right.
\ .
\fin
The explicit calculation of the SCE of a 2-dimensional half-shell
given in\equ{DiffNDS2} of Appendix~A is in agreement with the
estimate of\equ{CasNmD}. It is well known that the Casimir energy
is the same for a half-circle (i.e. interval) with Dirichlet,
respectively Neumann boundary conditions at the endpoints (since
there is no curvature for $d=1$ this also coincides with
ref.\cite{El94}). The Casimir energy due to a scalar on half an
Einstein universe has also been calculated explicitly\cite{KU80}
and was found to be just half of the Casimir energy for the full
Einstein universe irrespective of the boundary conditions (which
corresponds to $d=3$ in\equ{CasNmD}. Note that the fate of the
implicitly subtracted infinite vacuum energy proportional to the
surface area of the equator of the halved Einstein universe was
ignored in\cite{KU80}. Fulling has pointed out\cite{Fu89} that
(infinite) changes in the vacuum energy of the universe could be
absorbed in the cosmological constant. They also may be cancelled
by similar (infinite) contributions from other fields as in the
case of super-symmetry.

The following example of a $d$-dimensional parallelepiped shows
that the semiclassical evaluation of the Casimir energy of spaces
with boundaries that are not smooth and intersect on lower
dimensional manifolds can be even more involved.

\subsection{Semiclassical Casimir Energy of a $d$-dimensional
Parallelepiped}\label{parallele} The Casimir energy of a massless
scalar field confined to the interior of a parallelepiped with
dimensions $l_1\times l_2\times\dots\times l_d$ has previously
been obtained by Ambj{\"o}rn and Wolfram\cite{AW83}. They
considered Neumann, Dirichlet as well as periodic and
electromagnetic boundary conditions on the surface of the
parallelepiped. I here give a more geometrical interpretation of
some of their results in terms of periodic rays of a covering
space.

It suffices to consider the case where all the $l_i$ of the
parallelepiped are finite. The result of ref.\cite{AW83} for a
parallelepiped with some sides that are much longer than all
others is found by taking the appropriate limits. Lacking a more
concise notation, the SCE of a parallelepiped with dimensions
$l_1\times\dots\times l_d$ will be denoted by,
\bel{parallel}
{\cal E}_c(l_1,\dots,l_N;l_{N+1},\dots,l_{N+D};\,
l_{N+D+1},\dots,l_d)\ .
\fin
Here Neumann boundary conditions are satisfied on $0\leq N\leq d$
pairs of parallel hypersurfaces that are distances $l_1,\dots,l_N$
apart, Dirichlet boundary conditions are satisfied on $0\leq D\leq
d-N$ pairs of parallel hyper-surfaces that are distances
$l_{N+1},\dots,l_{N+D}$ apart and periodic boundary conditions are
assumed to hold on the remaining $0\leq d-N-D$ pairs of parallel
surfaces.

The symmetry of a parallelepiped implies that eigenfunctions
satisfying periodic boundary conditions on a pair of faces, are
even or odd under reflection of the parallelepiped about these
faces. This leads to the following relation between the Casimir
energies of scalar fields satisfying different boundary conditions
on the surfaces of $d$-dimensional parallelepipeds,
\bal{sumpiped}
&&{\cal
E}_c(l_2,\dots,l_{N};l_{N+1},\dots,l_{N+D};2l_1,l_{N+D+1},\dots,l_d)=\nonumber\\
&&\qquad{\cal
E}_c(l_1,\dots,l_N;l_{N+1},\dots,l_{N+D};l_{N+D+1},\dots,l_d)\\
&&\qquad+{\cal
E}_c(l_2,\dots,l_N;l_1,l_{N+1},\dots,l_{N+D};l_{N+D+1},\dots,l_d)\
.\nonumber
\ea
\equ{sumpiped} is a relation between the Casimir energies of a
scalar field on parallelepipeds where the periodic boundary
conditions on a pair of parallel surfaces are replaced by
Dirichlet or Neumann boundary conditions \emph{and} the distance
between the two surfaces is \emph{halfed}. The total volume of the
manifolds on the left- and right-hand sides of this equation thus
are the same. The subtracted terms of the spectral densities are
the same as well and the Casimir energies indeed are comparable.

The spectrum of a parallelepiped differs only in a zero frequency
mode for Neumann and Dirichlet conditions on a set of parallel
surfaces. This frequency does not depend on the separation of the
two surfaces. One therefore also has that,
\bal{diffpiped}
&&{\cal
E}_c(l_2,\dots,l_{N};l_{N+1},\dots,l_{N+D};l_{N+D+1},\dots,l_d)=\nonumber\\
&&\qquad {\cal
E}_c(l_1,\dots,l_N;l_{N+1},\dots,l_{N+D};l_{N+D+1},\dots,l_d)\\
&&\qquad-{\cal
E}_c(l_2,\dots,l_N;l_1,l_{N+1},\dots,l_{N+D};l_{N+D+1},\dots,l_d)\
.\nonumber
\ea
Note that the subtractions in the spectral density proportional to
the $d$-dimensional volume cancel on the right hand side and the
leading remaining subtraction is proportional to the volume of the
$(d-1)$-dimensional parallelepiped on the left hand side.

Combining\equ{sumpiped} and \equ{diffpiped} one obtains the
following recursive relation for the SCE of a parallelepiped,
\bal{recursive}
&&2{\cal
E}_c(l_1,\dots,l_N;l_{N+1},\dots,l_{N+D};l_{N+D+1},\dots,l_d)=\nonumber\\
&&\qquad\qquad {\cal E}_c(l_2,\dots,l_{N};l_{N+1},\dots,l_{N+D};
2l_1,l_{N+D+1},\dots,l_d)\nonumber\\ &&\qquad\qquad+ {\cal
E}_c(l_2,\dots,l_{N};l_{N+1},\dots,l_{N+D};l_{N+D+1},\dots,l_d)\nonumber\\
&&\\&&2{\cal E}_c(l_2,\dots,l_N;l_1,l_{N+1},\dots,l_{N+D};
l_{N+D+1},\dots,l_d)=\nonumber\\ &&\qquad\qquad {\cal
E}_c(l_2,\dots,l_{N};l_{N+1},\dots,l_{N+D};2l_1,l_{N+D+1},\dots,l_d)\nonumber\\
&&\qquad\qquad- {\cal
E}_c(l_2,\dots,l_{N};l_{N+1},\dots,l_{N+D};l_{N+D+1},\dots,l_d)\
.\nonumber
\ea

\equ{recursive} expresses the Casimir energy of a parallelepiped
in terms of Casimir energies of parallelepipeds with non-periodic
boundary conditions on fewer sets of parallel plates. Repeated
application of these relations thus gives the Casimir energy of a
parallelepiped with Neumann, Dirichlet and periodic boundary
conditions in terms of the Casimir energies of tori only. Thus the
Casimir energy of a three-dimensional parallelepiped with
Dirichlet boundary conditions on all six faces may be decomposed
as,
\bal{3Dpiped}
8{\cal E}_c(;l_1,l_2,l_3;)&=&4{\cal E}_c(;l_2,l_3;2l_1)-
4{\cal E}_c(;l_2,l_3;)\nonumber\\
&=&2{\cal E}_c(;l_3;2l_1,2l_2)-2{\cal E}_c(;l_3;2l_1)\nonumber\\&&
-2{\cal E}_c(;l_3;2 l_2)+2{\cal E}_c(;l_3;)\nonumber\\
&=&{\cal E}_c(;;2l_1,2l_2,2l_3)\\&& -{\cal E}_c(;;2l_1,2l_2)-{\cal
E}_c(;;2l_1,2l_3)-{\cal E}_c(;;2l_2,2l_3)\nonumber\\&& +{\cal
E}_c(;;2l_1)+{\cal E}_c(;;2l_2)+{\cal E}_c(;;2l_3)\ .\nonumber
\ea

The corrections proportional to the Casimir energies of lower
dimensional tori are due to periodic rays of the covering space
that lie on the boundaries of the original parallelepiped. Since
the subtraction for a torus is the Weyl-contribution proportional
to its surface, the subtractions for a general parallelepiped
include Weyl-terms proportional to the "area" of its
hyper-surfaces, "lengths" of the intersections of its
hyper-surfaces etc.. The Casimir energy of a general
parallelepiped thus can be compared with that of other systems of
the same volume, area of the boundary (with the same boundary
conditions), length of intersections of hyper-surfaces etc... The
simplest class of spaces that satisfy all these conditions are
$d$-dimensional generalizations of Power's box\cite{Po64} with a
fixed number of orthogonal but movable walls\footnote{Power's
original construction of a box with just one movable wall for the
Casimir force between two parallel plates has been reexamined and
extended in\cite{He05}. Since the surface areas of the two
parallelepipeds are linearly dependent on their volume,
considering a box with just one movable wall does not isolate the
surface dependence of the Casimir energy of a parallelepiped --
one has to consider more than one movable wall.}

Let us turn to the construction of the appropriate covering space
for a parallelepiped. The momentum is single valued only when
periodic boundary conditions hold on all pairs of hyper-surfaces.
$N+D=0$ and there is no need to introduce additional sheets. The
result is the same as for the $d$-dimensional torus in
section~\ref{torispheres}: the Casimir energy in this case is
negative for any dimension of the torus and any lengths, $l_i$, of
its cycles.

For $N+D>0$, the momentum is not single valued. Each pair of faces
with non-periodic boundary conditions requires a double covering
since the component of momentum that is perpendicular to these
surfaces can have either sign. One recursively constructs this
covering space as follows.

Consider first the pair of faces with coordinate $x_1=0$ and
$x_1=l_1$. The boundary condition on this pair is not periodic and
the first component of momentum (for rays of fixed energy)
therefore is double valued. It is single valued on a covering
space obtained by joining a second sheet of the original
parallelepiped to the first at the boundaries $x_1=0$ and
$x_1=l_1$ to form a cylinder-like covering space. Periodic rays
that reflect from the $x_1=0$ and $x_1=l_1$ faces of the original
parallelepiped pass smoothly through these borders from one sheet
to the other in the covering space. Although one must keep track
of the position of the original boundaries at $x_1=0$ and
$x_1=l_1$, the problem of constructing a covering space on which
momentum is unique has been reduced to that of a parallelepiped
with only $D+N-1$ pairs of hyper-surfaces on which non-periodic
boundary conditions hold. Note that the first dimension of this
covering space is now twice that of the original parallelepiped.

If $N+D>1$, the procedure is repeated with the pair of faces at
$x_2=0$ and $x_2=l_2$ of this double cover of the original
parallelepiped. This results in a covering parallelepiped with
non-periodic conditions in just $N+D-2$ dimensions. The length of
this covering parallelepiped in the second dimension is again
twice that of the original parallelepiped. The process ends with
the pair of faces at $x_{D+N}=0$ and $x_{D+N}=l_{D+N}$.

The covering space of a general $d$-dimensional parallelepiped
thus is a $d$-dimensional torus with cycles
$2l_1,\dots,2l_{D+N},l_{D+N+1},\dots,l_d$. The original
parallelepiped is covered $2^{D+N}$ times. The $2(D+N)$
$(d-1)$-dimensional hyper-surfaces with non-periodic boundary
conditions of the original parallelepiped now are
$(d-1)$-dimensional interfaces between sheets of this covering
space at $x_1=0,x_1=l_1,x_2=0,\dots,x_{D+N}=0$ and
$x_{D+N}=l_{D+N}$.

Since a general periodic ray crosses pairs of boundary surfaces,
$\beta_\gamma=0$ for periodic- and Neumann, respectively Dirichlet
boundary conditions on opposing pairs of
hyper-surfaces\footnote{$\beta_\gamma ={\rm even}\geq 2$ if the
scalar field satisfies Neumann's condition on one, but Dirichlet's
condition on the other of a pair of parallel hyper-surfaces. Such
asymmetric boundary conditions on a pair of parallel surfaces can
give a change in sign of the semi-classical Casimir
energy\cite{MS98,Hu97}. It is due to a phase loss of $\pi$ for
some dominant primitive periodic rays and can be explained along
the lines used for manifolds without boundary\cite{MS98}.}.

On this toroidal covering space there are periodic rays that lie
entirely within the $d-1$-dimensional hyper-surfaces that are
projected onto the boundaries of the original parallelepiped. The
contribution to the Casimir energy due to these rays depends on
the imposed boundary conditions. The hyper-surfaces of a
parallelepiped in addition intersect on lower dimensional
hyper-edges that also contain periodic rays of the covering space.

The corrections due to rays on these lower-dimensional tori are
clearly visible in\equ{recursive} and\equ{3Dpiped}. They are in
one-to-one correspondence with the boundary surfaces, edges, etc.
of the parallelepiped. The contribution of rays on the
lower-dimensional surfaces has to be subtracted for Dirichlet
boundary conditions because the field vanishes on this surface of
the parallelepiped in this case.  That the lower-dimensional
correction in\equ{recursive} is added for Neumann conditions then
follows from the reflection symmetry of the parallelepiped, which
implies \equ{sumpiped}.

The sign of the Casimir energy of a parallelepiped in general
therefore is determined by the boundary conditions on its
surfaces. One can argue that the Casimir energy due to periodic
rays of a boundary surface is always negative when this surface is
embedded in a higher-dimensional space of vanishing curvature:
there are no caustics to contend with and the energy surface has
no zero modes.

Whether the Casimir energy due to periodic rays on such a boundary
surface has to be subtracted or added depends on the imposed
condition. The sign of the overall Casimir energy of a
parallelepiped therefore depends on the relative magnitude of
boundary contributions with opposite sign. Which sign prevails in
general will depend on the actual dimensions of the
parallelepiped. The previous analysis nevertheless allows for a
few general statements about the sign of the Casimir energy of a
parallelepiped:

\begin{itemize}
    \item Since contributions from periodic rays on lower dimensional boundary surfaces are always
negative, the sign of the SCE of a parallelepiped with only
Neumann and periodic boundary conditions is negative in any
dimension.
    \item Replacing Dirichlet- by Neumann- boundary conditions on a pair of parallel surfaces
decreases the Casimir energy of the parallelepiped.
    \item If one dimension of the parallelepiped is much smaller than all others,
as for two parallel infinite hyper-planes, the difference in
Casimir energy for Neumann and Dirichlet boundary conditions tends
to vanish (since all contributions to the Casimir energy due to
rays on lower dimensional surfaces become negligible).
\end{itemize}

\section{Conclusion}
By defining the SCE through the required subtraction $\rho_0(E)$
in the spectral density of a system one describes a class of
systems whose vacuum energies can be compared. By definition the
subtracted spectral density $\tilde\rho(E)$ of\equ{rho} gives a
finite Casimir energy. It is approximated semiclassically by
contributions due to classical periodic rays. For a massless
scalar field on $d$-dimensional spheres and tori as well as on
related spaces with boundary such as half-spheres and
parallelepipeds, this SCE coincides with other definitions
whenever the subtractions in the spectral density are the same for
a non-trivial class of systems(see Appendix~A).

[One may argue that certain systems, such wedges formed by two
semi-infinite planes joined at the common edge\cite{MT97}, do not
have classical periodic rays and that this semiclassical approach
may thus be rather limited in scope. However, the implicit
subtraction is not universal in this case, and depends on the
opening angle of the wedge in a non-linear fashion\cite{BD04} that
prohibits comparisons between systems with wedges of different
opening angle. The extracted finite Casimir energy of a particular
wedge in this case is of little physical significance and for
instance does \emph{not} determine the torque between the two
plates of the wedge.]

The geometrical description of the SCE in terms of periodic rays
gives insights into qualitative features of this part of the
vacuum energy that otherwise are rather mysterious. The need for
an explanation of the sign of Casimir energies was very nicely
formulated by J.~Sucher:``Understanding the signs is a sign of
understanding''\cite{Hu97}. I have here presented some evidence
that the sign of the SCE depends critically on optical properties
of "important" (short) periodic rays.

The semiclassical contribution to the Casimir energy due to an
isolated periodic ray has a definite sign (see \equ{CasimirE}).
The contribution due to a class of periodic rays of an integrable
system also is of a definite sign (see \equ{Casintfin}. For
isolated periodic rays, the sign is determined by the winding
number $\sigma_\gamma$ of their stable and unstable
manifolds\cite{CRL90}, whereas it essentially is given by the
Keller-Maslov index $\beta_{{\bf n}}$ of a class of periodic rays
in integrable systems. [When the Hessian of \equ{hess} is not
positive definite, the sign of the contribution of a class of rays
is more appropriately given by\equ{identify}.]

The semiclassical expressions of\equ{CasimirE} and\equ{Casintfin}
suggest that the Casimir energy very often is dominated by the
shortest periodic rays or class of periodic rays. Arguing that the
contribution of the shortest periodic rays is continuous under
small deformations of the manifold one can estimate the sign of
the Casimir energy of integrable systems in two ways: either by
direct computation of the index in\equ{identify}, or by slight
deformation of the integrable system and computation of the index
of\equ{sigma} for the shortest periodic rays. Continuity under
deformations also explains the pattern of signs for the Casimir
energies of a massless scalar on spheres and of tori of various
dimensions as well as changes in sign when spheres are strongly
deformed. Sign changes of the SCE in these cases are accompanied
by a change the number of conjugate points of the shortest rays.

The sign of the SCE is harder to determine for spaces with
boundary. The shortest periodic rays again dominate, but the sign
(and magnitude) of contributions due to periodic rays that lie
within the boundary depends on the boundary conditions. Such
boundary rays can be among the shortest periodic rays and in some
cases dominate the Casimir energy.

Only $d$-dimensional half-spheres and general parallelepipeds with
Neumann and Dirichlet boundary conditions were considered.
However, the general arguments remain valid in more realistic
situations with, for instance, electromagnetic fields. The sign of
the Casimir energy of spherical- and cylindrical- cavities with
idealized metallic boundary conditions can apparently be
understood in terms of the phases of the shortest periodic
rays\cite{future}.

Determining the sign of a SCE in this sense is reduced to a
problem of geometrical optics. For certain simple manifolds, such
as tori and spheres, one finds that \emph{all} contributions due
to periodic rays are of the same sign. The sign of the SCE is
unambiguous in this case. The fact that classical dynamics of
periodic trajectories seems to determine the sign of the Casimir
energy raises the intriguing question whether the sign is a
topological characteristic of the phase space. In general, and in
particular for manifolds with boundaries, periodic rays of
comparable length contribute to the SCE with opposite sign. It
then depends on the boundary conditions and metric characteristics
of the space whether the SCE is positive or negative.

{\bf Acknowledgements:} I am greatly indebted to Larry Spruch for
suggesting this investigation. This work would not have been
possible without his encouragement and support and it benefitted
greatly from numerous discussions. I also would like to thank
S.~A.~Fulling for a critical review of an earlier version of the
manuscript and for the invitation to a superbly organized and very
interesting workshop on semiclassical approximation and vacuum
energy.

\appendix
\section{Casimir Energies and Curvature Corrections: Spheres and Half-Shells}
I here explicitly compute the Casimir energies due to a scalar
field for low-dimensional spheres $S_1$, $S_2$, $S_3$ and $S_4$ as
well as for the two-dimensional half-shell with Neumann,
respectively Dirichlet boundary conditions. Of special interest
are the associated subtractions in the spectral density that
render the Casimir energies finite. As explained in the main text,
these subtractions determine classes of systems with finite vacuum
energy differences. The finite Casimir energy is given (exactly)
by contributions due to periodic rays whenever the subtractions
are universal.

\subsection{The circle}
The Casimir energy of a scalar field on $S_1$ probably is the most
transparent example. The curvature vanishes and the energy
spectrum of a massless field on $S_1$ clearly is $E_l=l\hbar c/R$
for integer $l\geq 0$. For $l>0$ the energy eigenvalues are 2-fold
degenerate. The vacuum energy of a massless scalar on $S_1$ thus
is formally given by,
\bel{ES1}
E_{\rm vac}(S_1)=\frac{\hbar c}{2 R}\sum_{l=1}^\infty 2 l\ .
\fin
One may regularize this divergent expression in many ways. One of
the more popular is by analytic continuation in the exponent $s$
of the for $s<-1$ manifestly convergent sum
$\zeta(-s)=\sum_1^\infty l^s$. This method, known as zeta function
regularization, has been claimed\cite{El94} to be not just the
most elegant, but also the only rigorous mathematical definition
of the sum in\equ{ES1}. That may be true but gives appreciably
little insight into the physical significance of the finite
Casimir energy one obtains, which for a circle is $-\frac{\hbar
c}{12 R}$.

Let us subtract from the spectral density
$\rho_{S_1}(E;S_1)=\sum_{l=0}^\infty 2\delta(E-\hbar c l/R)$ the
smooth Weyl density $\rho_0(E)=\frac{2\pi R}{\pi\hbar c}$. The SCE
of a circle then is \emph{defined} to be,
\bal{ES1a}
{\cal E}_c(S_1)&=&{\frac 1 2}\int_0^\infty
\left[\rho(E;S_1)-\rho_0(E)\right]E dE\nonumber\\
&=&{\frac 1 2}\int_0^\infty
\left[\sum_{l=0}^\infty 2\delta(E-\hbar c l/R)-\frac{2 R}{\hbar c}\right]E dE\nonumber\\
&=&-\frac{1}{\pi}{\rm Im}\frac{2\pi R}{i\hbar
c}\sum_{n=1}^\infty\int_0^\infty E dE\; e^{i n 2\pi R E/(\hbar
c)}\nonumber\\
&=&-\frac{\hbar c}{2\pi^2
R}\sum_{n=1}^\infty\frac{1}{n^2}=-\frac{\hbar c}{12 R}\ .
\ea
The second line of\equ{ES1a} expresses the subtracted spectral
density in terms of the semiclassical contribution due to periodic
rays of length $2\pi R n,\ n>0$.  The final answer coincides with
that of zeta function regularization, but one now explicitly
\emph{knows} the implicit subtraction in the spectral density
required to obtain it. The subtracted Weyl contribution to the
spectral density is proportional to the circumference of the
circle. The difference in Casimir energies of two circles thus is
\emph{not} the difference in their vacuum energy. It in fact would
cost an infinite amount of energy to change the radius of the
circle by a finite amount.

The subtraction nevertheless is universal in the sense that the
(so defined) finite Casimir energy reproduces all derivatives of
the vacuum energy with respect to the radius of $2^{\rm nd}$ and
higher order. The difference in total Casimir energy for instance
gives the energy required to change the \emph{relative} radii of
disjoint circles while keeping their sum constant\footnote{The
situation is unstable due to the minus sign of the Casimir energy
in\equ{ES1a}. It implies that the energy is minimized by shrinking
all but one of the circles to a point.}.

\subsection{The Two-Sphere}
The $2$-dimensional sphere $S_2$ is a prototypical manifold with
positive curvature. The Casimir energy of a scalar field on $S_2$
depends on how the scalar couples to the curvature. It is
known\cite{Gu90,Sc81,Fu89}, and we shall see, that this coupling
to the curvature is of a particular strength for a massless field.
However, to better compare with ref.\cite{El94}, let us for the
moment ignore the coupling to the curvature.

The eigenfunctions of the Laplace-Beltrami operator on a
two-sphere are the spherical harmonics with $(2l+1)$-fold
degenerate eigenvalues. Without coupling to the gaussian curvature
of $S_2$, the energy eigenvalues of a scalar field are $E_l=\hbar
c\sqrt{l(l+1)}/R$. The vacuum energy of a two-sphere of radius $R$
in this case is formally given by the divergent sum
\bel{ES2}
E_{\rm vac}(S_2)=\frac{\hbar c}{2 R}\sum_{l=0}^\infty (2
l+1)\sqrt{l(l+1)}.
\fin
Zeta function regularization (ZR) of\equ{ES2} gives\cite{El94} the
\emph{negative} value,
\bel{S2z}
{\cal E}_{ZR}(S_2)=\lim_{s\rightarrow 1/2}\frac{\hbar c}{2
R}\sum_{l=0}^\infty (2 l+1)[l(l+1)]^{s}=-0.132548\frac{\hbar c}{R}
\fin
for this Casimir energy. A physically perhaps more transparent
treatment of the sum in\equ{ES2} by heat kernel regularization
(HK) gives,
\bel{ES2b}
{\cal E}_{HK}(S_2)=\lim_{\eps\rightarrow 0_+}\frac{\hbar c}{2
R}\left\{\sum_{l=0}^\infty (2 l+1)\sqrt{l(l+1)}e^{-\eps l(l+1)}
-\frac{\sqrt{\pi}}{2\eps^{3/2}}\right\}= -0.132548\frac{\hbar
c}{R}.
\fin
Subtracting the integral over $l\in [0,\infty]$ from the sum over
angular momenta thus leads to the same finite result for the
Casimir energy as zeta function regularization. However, no
obvious \emph{physical} principle apart from finiteness of the
result seems to dictate the particular subtraction in\equ{ES2b}.
[For a justification and the relation to subtractions in other
regularization schemes see ref.\cite{Fu03}].

A straightforward physical interpretation of the subtraction is
possible in the semiclassical treatment: one again subtracts a
smooth "classical" part of the spectral density. $\rho_0(E)$
should not depend on the detailed shape of the surface. For a
2-dimensional manifold without boundaries, one may subtract the
classical Weyl contribution to the spectral density proportional
to the area $A$ of the manifold, $\rho_0(E)=AE/(2\pi(\hbar c)^2)$.
The subtracted spectral density is,
\bel{S2rs}
\tilde\rho(E)=\rho(E)-\rho_0(E)=\left[\sum_{l=0}^\infty(2
l+1)\delta\left( E-\hbar c\sqrt{l(l+1)}/R\right)\right]  - \frac{2
R^2E}{(\hbar c)^2}.
\fin
$\tilde\rho(E)$ can be expressed in terms of semiclassical
contributions from periodic rays\cite{Gu90, St87},
\bel{S2rc}
\tilde\rho(E)=-\frac{1}{\pi}{\rm Im}\,\frac{4\pi R^2 E}{i (\hbar
c)^2}\sum_{n=1}^\infty  \exp\left[i n\left(\frac{2\pi R}{\hbar
c}\sqrt{E^2+\left(\frac{\hbar c}{2 R}\right)^2}
-\pi\right)\right]\ .
\fin
One verifies that\equ{S2rc} is equivalent to\equ{S2rs} by
recognizing that apart from a $n=0$ term, the real part of the sum
over $n$ in\equ{S2rc} is a periodic $\delta$-distribution. The sum
in\equ{S2rc} on the other hand can be interpreted as due to
contributions from classical periodic rays that wind $n$ times
about a geodesic of the two-sphere with a total length $L_n=2\pi R
n$.

The Casimir energy corresponding to the spectral density
$\tilde\rho(E)$ of\equ{S2rc} is,
\bal{S2E}
{\cal E}_c(S_2)&=&{\frac1 2}\int_0^\infty E\tilde\rho(E) dE\nonumber\\
&=& \frac{\hbar c}{4 R}{\rm Re}\, e^{3i\pi/4}
\sum_{n=1}^\infty\int_0^\infty
(1+i\xi)\sqrt{\xi(2+i\xi)}\;e^{-n\pi\xi}d\xi\nonumber\\
&=&\frac{\hbar c}{4 R}{\rm Re} (i-1) \int_0^\infty
\frac{(1+i\xi)\sqrt{\xi(1+i\xi/2)}}{e^{\pi\xi}-1} d\xi\nonumber\\
&=& -0.132548\dots \frac{\hbar c}{R},
\ea
where the last number was obtained by numerical evaluation of the
integral\footnote{Substitution of the variable $i\xi(E)=
\sqrt{1+\left(\frac{2RE}{\hbar c}\right)^2 }-1$ and a clockwise
rotation of the integration contour by $90^o$ gives the integral
that converges exponentially and is numerically well behaved.}.
The SCE thus coincides with that of zeta function- and heat
kernel- regularization. The Casimir energy of a spherical shell is
completely described in terms of its classical periodic rays.
Moreover, this procedure gives a transparent physical meaning to
the subtraction: the spectral density $\tilde\rho(E)$ is the
difference to the universal Weyl term in the spectral density for
2-dimensional surfaces of the same area. The Casimir energy thus
does \emph{not} give the energy required to change the radius of
$S_2$: the subtracted (divergent) term proportional to the surface
area of the sphere is far more important for this energy
difference\cite{Ba01}. The Casimir energy on the other hand
\emph{does} allow the computation of the energy required to
\emph{deform} the sphere into another smooth manifold without
boundary of the \emph{same} total area. It thus for instance makes
sense to speak of the energy required to change a two-sphere to a
discus or an elongated ellipsoid of the \emph{same} area. The
energy required for such deformations is finite.

The effective action of a periodic ray of arc length $L_n=2\pi R
n$ in\equ{S2rc} is,
\bel{S2ac}
S_n =\oint {\bf p}\cdot {\bf dx}=p(E)L_n,
\fin
with
\bel{S2p}
p^2(E)=(E/c)^2+\hbar^2/(4R^2)\ .
\fin
\equ{S2p} is not the dispersion relation one expects for a
massless particle. $p^2(E)$ differs from $(E/c)^2$ by a term of
order $\hbar^2$ that is proportional to the Gaussian curvature
$\kappa=1/R^2$ of $S_2$. The curvature of the manifold results in
a dispersion relation that would corresponds to a tachyon with
velocity $v(E)=(dp/dE)^{-1}=c \sqrt{1+(\hbar c/(2 R E))^2}>c$.
This can also be seen by rewriting,
\bel{S2we}
E_l^2=\frac{(\hbar c)^2}{R^2}l(l+1)=\left(\frac{\hbar c
(l+1/2)}{R}\right)^2 -\left(\frac{\hbar c}{2 R}\right)^2.
\fin
Interpreting $\hbar (l+1/2)/R$ as the eigenvalue of the momentum
operator for a field satisfying anti-periodic boundary conditions
on a circle of radius $R$, this dispersion clearly is tachyonic:
the effective mass squared is negative $m^2=-(\hbar/2 c R)^2$. The
fact that the effective action in\equ{S2ac} does not vanish for
$E=0$ is another indication that the spectrum of the
Laplace-Beltrami operator on curved spaces is not the spectrum of
a massless scalar field.

A particle is massless if its dispersion is $p(E)=E/c$. The
generalization of the wave equation for $d$-dimensional Riemannian
manifolds ${\cal M}$ with metric $g_{ij}$ and Gaussian curvature
$\kappa$ includes a coupling to the curvature. The appropriate
wave equation for a $d+1$-dimensional (ultrastatic) space-time
with curvature is $\Delta_d\phi=c^{-2}\partial^2_t\phi$, where
\bel{Laplace}
\Delta_d=(1/\sqrt{g})\frac{\partial}{\partial
x^i}\sqrt{g}g^{ij}\frac{\partial}{\partial x^j}-\frac{(d-1)^2}{4
}\kappa\ .
\fin
This curvature correction to the Laplace-Beltrami operator of
${\cal M}$ also arises naturally from the measure of the path
integral and is consistent with the required short-time behavior
of the Feynman propagator\cite{Sc81}. The particular strength of
the coupling to the curvature in\equ{Laplace} preserves conformal
invariance\footnote{$\kappa$ in\equ{Laplace} is the Gaussian
curvature of ${\cal M}$ rather than the Ricci curvature scalar
${\cal R}$ of the space-time ${\bf R}\times {\cal M}$ considered
in\cite{Fu89}. For an ultrastatic space-time the two curvatures
are related by ${\cal R}=\kappa(d-1)/d$ and the particular
coupling strength in\equ{Laplace} corresponds to the conformal
coupling $\xi=((d+1)-2)/(4((d+1)-1))$ discussed in\cite{Fu89}. }
of the wave equation\cite{Fu89}. For a $d$-dimensional sphere
$S_d$ of radius $R$, $\kappa=1/R^2$ and the eigenvalues of
$-\Delta_d$ defined in\equ{Laplace} are $(l+(d-1)/2)^2/R^2$ for
integer $l\geq 0$. The Casimir energy due to a \emph{massless}
scalar field on a two-sphere thus is,
\bal{s2}
{\cal E}_c(S_2;m=0)&=&\int_0^\infty \frac{E dE}{2}
\left[\sum_{l=0}^\infty (2l+1)\delta(E-\hbar c(l+1/2)/R)
- \frac{2 R^2E}{(\hbar c)^2}\right]\nonumber\\
&=& -\frac{1}{2\pi}{\rm Im}\,\frac{4\pi R^2 }{i (\hbar
c)^2}\sum_{n=1}^\infty \int_0^\infty E^2 dE \exp\left[i
n\left(\frac{2\pi R}{\hbar c}E -\pi\right)\right]\nonumber\\
&=&{\rm Im}\,\frac{\hbar c }{4\pi^3
R}\sum_{n=1}^\infty\frac{(-1)^n}{n^3}=0\ .
\ea
That the Casimir energy of a massless scalar field vanishes for a
two-sphere agrees with the arguments presented in
Sections~\ref{torispheres} and~\ref{integrablesys}. Zeta function-
and heat kernel-regularization also give a vanishing Casimir
energy when the coupling of a massless scalar to the curvature is
included. Note that the average spectral density of a
2-dimensional manifold of the same area again was subtracted to
obtain the finite answer of\equ{s2}. Although the second line
of\equ{s2} is a sum over contributions due to periodic rays of the
two-sphere, it does not have the form of\equ{response}. The
amplitude in particular is proportional to the energy here. As
discussed in Section~\ref{integrablesys}, free motion on a
two-sphere is integrable and the corresponding periodic rays are
not isolated. As shown in Appendix~B, a semiclassical treatment
along the lines of ref.\cite{BT76} gives the semiclassical
response function of\equ{s2} without reference to the exact
quantum mechanical spectrum.

Let us now turn to a manifold with boundary and consider a
two-dimensional half-sphere $S_2/2$ with Neumann (N), respectively
Dirichlet (D) boundary conditions on the equatorial
circle\footnote{A more general treatment of the Selberg trace
formula for two-dimensional manifolds with boundaries is given
in\cite{BS93}. However, the particular example of a massless
scalar on a half-sphere, does not appear to satisfy the
restriction that $h(p)=e^{-|p| t}$ be analytic on the strip $|{\rm
Im}\, p|<1/2+\eps$.} Due to the symmetry of the two-sphere upon
reflection about its equator, eigenfunctions of the
Laplace-Beltrami operator either satisfy Neumann or Dirichlet
conditions on the equator and thus ${\cal E}(S_2/2;N)+{\cal
E}(S_2/2;D)={\cal E}(S_2)=0$. Since one more mode satisfies
Neumann's boundary condition than  Dirichlet's for every energy
eigenvalue, one has for the Casimir energy of a half-sphere,
\bal{DiffNDS2}
{\cal E}_c(S_2/2;{\scriptstyle {N \atop D}})&=&\pm\int_0^\infty
\frac{ E dE}{4}\left[\sum_{l=0}^\infty\delta(E-\hbar c(l+1/2)/R)
- \frac{R}{\hbar c}\right]\nonumber\\
&=&\mp\frac{1}{\pi}{\rm Im} \frac{2\pi R}{i\hbar c}
\sum_{n=1}^\infty\int_0^\infty \frac{ E dE}{4} \exp\left[i
n\left(\frac{2\pi R}{\hbar c}E -\pi\right)\right]\nonumber\\
&=&\mp\frac{\hbar c}{8\pi^2 R}\sum_{n=1}^\infty
\frac{(-1)^n}{n^2}={\pm}\frac{\hbar c}{96 R}\ .
\ea
This again agrees with zeta function regularization for this case.
Note that an additional universal subtraction is necessary. It is
a Weyl contribution to the spectral density proportional to the
\emph{length} of the boundary. Due to the subtraction of terms
proportional to the area of the surface and proportional to the
length of the boundary, the Casimir energy of a massless scalar on
the half-sphere can be compared to the Casimir energy on another
smooth 2-dimensional manifold of the \emph{same area} and with a
smooth boundary of the \emph{same length} only.

Note that the necessary subtraction for a half-sphere with
boundary is \emph{not} universal if the curvature correction to
the Laplace-Beltrami operator in\equ{Laplace} is ignored. In this
case one obtains the following difference in Casimir energies,
\bal{DiffNDS2L}
{\cal E}(S_2/2;N)&-&{\cal E}(S_2/2;D)=\nonumber\\&=&\int_0^\infty
\frac{ E dE}{2}\left[\sum_{l=0}^\infty\delta(E-\hbar
c\sqrt{l(l+1)}/R) -
\frac{R}{\hbar c}\frac{E}{\sqrt{E^2+\left(\frac{\hbar c}{2 R}\right)^2}}\right]\nonumber\\
&=&-\frac{1}{\pi}{\rm Im} \frac{2\pi R}{i\hbar c}
\sum_{n=1}^\infty\int_0^\infty \frac{E dE}{2}\frac{E}{
\sqrt{E^2+\left(\frac{\hbar c}{2 R}\right)^2}}\ e^{i
n\left(\frac{2\pi R}{\hbar c}\sqrt{E^2+\left(\frac{\hbar c}{2
R}\right)^2} -\pi\right)}\nonumber \\
&=&\frac{\hbar c}{4 R}{\rm Re} (i-1)\int_0^\infty d\xi
\frac{\sqrt{\xi(1+i \xi/2)}}{e^{\pi\xi}-1}
=-0.110687\dots\frac{\hbar c}{R}.
\ea
The subtracted spectral density $\rho_0(E)$ in\equ{DiffNDS2L} is
again proportional to the length of the boundary, but it now
depends on the Gaussian curvature $\kappa=1/R^2$ as well.
Expanding for small values of the curvature
$E/\sqrt{E^2+\left(\frac{\hbar c}{2 R}\right)^2}\sim
1-\kappa(\hbar c)^2/(8E^2)$ shows the presence of a (logarithmic)
divergent contribution to the vacuum energy that is proportional
to $1/R$. No derivative of the vacuum energy with respect to the
radius of the half-shell is finite and the subtracted (finite)
Casimir energy of\equ{DiffNDS2L} is of no physical significance.

In zeta function regularization the logarithmic divergence
manifests itself as a pole. The principal value prescription
ignores this pole contribution to the Casimir energy to produce a
finite value\cite{El94} of $-0.166080 \hbar c/R$ for the above
difference of the Casimir energies of half-shells with Neumann and
Dirichlet boundary conditions. Since the implicit subtraction is
not universal, it should not surprise that the value ascribed by
principal value prescription differs from the one
in\equ{DiffNDS2L} -- obtained by prescribing a particular
subtraction of the spectral density. The finite part of the
Casimir energy of this system is quite arbitrary since its vacuum
energy cannot be compared to that of any other system anyway. The
appearance of a pole in zeta function regularization in this sense
indicates that the associated implicit subtraction cannot be
universal and that the finite value obtained by some prescription
is not very meaningful.

If one is of the opinion that a Casimir energy due to a physical
field should enjoy some degree of universality, small deformations
of a smooth manifold should not require an infinite amount of
energy. The lack of universality of the Casimir energy on a
half-shell in this sense is another indication that the spectrum
of the Laplace-Beltrami operator is not that of a physical
particle. This favors\equ{Laplace} as the appropriate
generalization of the wave operator to manifolds with curvature.
The argument perhaps is more convincing if one notes that even the
Casimir energy of a scalar on $S_3$ (a manifold without boundary
that could serve as a model of the spatial part of our universe)
is not universal\cite{El94} without curvature correction.

\subsection{$S_3$ and $S_4$}
For comparison with the results of ref.\cite{El94} and in support
of the arguments for the sign of the Casimir energy of a massless
scalar field on $d$-dimensional spheres of
Sections~\ref{torispheres} and~\ref{integrablesys}, I here give
the universal Casimir energies of a massless scalar field on $d=3$
and $d=4$-dimensional spheres. I have included the appropriate
coupling to the curvature of\equ{Laplace} and the results
therefore differ from those without curvature correction
in\cite{El94}. The difference is not minor: the Casimir energy of
a scalar on $S_3$ in particular is not universal and devoid of
physical implications without coupling to the curvature. In zeta
function regularization this is indicated by the appearance of a
pole\cite{El94}.

The Casimir energies due to a scalar on $S_3$ and $S_4$ can be
obtained in much the same manner as for the two-sphere -- the main
difference is the degeneracy of the spectrum of the
Laplace-Beltrami operator in\equ{Laplace}. It is $(l+1)^2$ for
$S_3$ and $(l+1)(l+2)(2l+3)/6$ for $S_4$. As can be seen
from\equ{Laplace}, the spectrum of a massless field with the
appropriate coupling to the curvature is $E_l=\hbar c (l+1)/R $ on
$S_3$ and $E_l=\hbar c (l+3/2)/R$ on $S_4$. One then finds,
\bal{S3S4}
{\cal E}_c(S_3;m=0)&=&\frac{\hbar c}{240 R}=+0.0041666\dots
\frac{\hbar c}{R}\nonumber\\
{\cal E}_c(S_4;m=0)&=&0\ .
\ea
These results again are in agreement with those of zeta function
regularization for this case. There are no logarithmic divergent
subtractions and no poles appear in zeta function regularization
when the appropriate curvature correction is taken into account.
For $S_4$ the subtraction of the Weyl contribution proportional to
the 4-volume of $S_4$ is not sufficient to obtain a finite Casimir
energy. An additional smooth spectral density proportional to
$R^2$ must be subtracted and,
\bel{rho4}
\rho_0(E;S_4)=\frac{\pi R^4 E^3}{3(\hbar
c)^4}\left[1-\left(\frac{\hbar c}{2 R E}\right)^2\right]\ .
\fin
The additional term in the smooth part of the spectral density is
proportional to the \emph{integrated} curvature of the manifold.
One thus can compare the (vanishing) Casimir energy of a massless
scalar field on $S_4$ with that of other $4$-dimensional manifolds
without boundary of the \emph{same 4-volume and average
curvature}.

\section{Semiclassical Derivation of the Spectral Density of a Scalar on a Two-Sphere}
For completeness I here give the semiclassical calculation of the
Casimir energy ${\cal E}_c(S_2)$ of\equ{s2} for a massless scalar
particle on a two-sphere without reference to the quantum
mechanical spectrum. Although a somewhat trivial example of the
more general asymptotic expansion for elliptic self-adjoint
pseudo-differential operators on Riemannian
manifolds\cite{Ch74,DG75}, it does illustrate some of the basic
ideas at an elementary level and illustrates the general procedure
of section~\ref{integrablesys}.

The system has two constants of motion in involution and is thus
integrable. One of these can be taken to be the square or
magnitude of the momentum $|p|$, the other the angular momentum
$l_z$. The corresponding action variables for a two-sphere of
radius $R$ are $I_1=R |p|\geq 0$ and $I_2=l_z$ with $|I_2|\leq
I_1$. The hamiltonian for a free massless particle on the
two-sphere is $H(I_1,I_2)=c |p|=c I_1 /R$. Following
ref.\cite{BT76,Gu90}, the oscillating part of the semiclassical
spectral density is,
\bel{semiS2}
\tilde\rho(E;S_2)=\frac{1}{\hbar^2}\sum_{n,m=-\infty}^{\ \ \infty\
\prime} \int_0^\infty dI_1\int_{-I_1}^{I_1} dI_2
\delta(E-H(I_1,I_2))\ e^{2\pi i[( n I_1+m I_2)/\hbar-n/2]}
\fin
The integration domain of the action variables in\equ{semiS2} is
due to classical considerations only and the primed sum
in\equ{semiS2} here implies that the (classical) contribution with
$m=n=0$ is to be omitted. Note that the $m=n=0$-term is just the
$\rho_0(E)$ that is subtracted from the full spectral density. The
spectral density evidently is given in terms of contributions from
periodic trajectories that wind $(n,m)$ times about the (in this
case two-dimensional) invariant tori of constant energy. The phase
retardation by $n\pi$ of the family of trajectories that wind
$n$-times about the whole two-sphere is related to the uniqueness
of the quantum mechanical wave function\cite{Ke58} and might not
be expected classically. Keller's construction in
section~\ref{coveringspaces} shows that this phase loss is due to
the two caustics of first order formed by families of periodic
rays of the same inclination\footnote{For fixed energy the
inclination of a periodic orbit is determined by $l_z=I_2$ on
$S_2$}. Inserting\equ{semiS2} in\equ{Cascdef} and performing the
integral over $E$ one obtains,
\bel{ES2semi}
{\cal E}_c(S_2) = \frac{1}{2\hbar^2}\sum_{n,m=-\infty}^{\ \
\infty\ \prime} \int_0^\infty dI_1 \int_{-I_1}^{I_1} dI_2
H(I_1,I_2)\ e^{2\pi i[( n I_1+m I_2)/\hbar-n/2]}\ .
\fin
Although a direct consequence of\equ{semiS2}, it perhaps is
remarkable that the SCE of an integrable system is expressible in
terms of the Fourier-transform of the classical hamiltonian with
respect to action variables. Since the hamiltonian $H(I_1,I_2)= c
I_1/R$ does not depend on $I_2$ and one is interested in the term
of order $\hbar$ only, one can perform the integral over $I_2$ in
saddle point approximation. This gives,
\bel{saddle}
\int_{-I_1}^{I_1} dI_2\ e^{2\pi i m I_2/\hbar}=\hbar
\int_{-I_1/\hbar}^{I_1/\hbar} dx\ e^{2\pi i m x}=2 I_1
\delta_{m0}+ {\cal O}(\hbar)\ .
\fin
The only sizable contribution to the $I_2$ integral thus is from
the $m=0$ term of the sum. Inserting\equ{saddle} in\equ{ES2semi}
one finally obtains for the SCE due to a massless scalar on $S_2$,
\bal{EsemiS2}
{\cal E}_c(S_2) &=& \frac{c}{R \hbar^2} \sum_{n\neq 0}
(-1)^n\int_0^\infty dI_1 I_1^2\ e^{2\pi i n I_1/\hbar}\nonumber\\
&=&\frac{-i\hbar c}{(4\pi^3) R} \sum_{n\neq 0}
\frac{(-1)^n}{n^3}=0\ .
\ea
The integral over $I_1$ in\equ{EsemiS2} coincides with that over
the energy in\equ{s2} if one recalls that $I_1=E R/c$. Note that
higher moments of the semiclassical spectral density do not
vanish. One in particular finds that the second moment of the
semiclassical spectral density is negative,
\bel{S2E2}
\left< E^2\right>_c(S_2)=12\left(\frac{\hbar c}{4 \pi^2
R}\right)^2 \sum_{n\neq 0} \frac{(-1)^n}{n^4}= - \frac{7}{480}
\left(\frac{\hbar c}{R}\right)^2\ .
\fin
The contribution of periodic rays to the spectral density of a
massless scalar thus cannot be positive semi-definite for $S_2$.
Upon subtracting the "classical" Weyl contribution to the spectral
density there in fact is no reason why the remainder must be
positive. The negative sign in\equ{S2E2} can be traced to the
phase loss of $\pi$ for every revolution of a periodic ray and
thus ultimately to the Keller-Maslov  index of a class of periodic
rays.

\end{document}